\newcommand{\ignore}[1]{}

\documentclass[9pt, noacm, sigconf]{acmart}
\usepackage{url}
\usepackage{multirow}
\usepackage{subfigure}
\usepackage{epsfig}
\usepackage{graphicx}
\usepackage{amsmath}
\usepackage{color}
\usepackage{setspace}
\usepackage{hhline}   
\usepackage{algorithm}
\usepackage{algpseudocode}
\usepackage{flushend}
\usepackage{url}
\usepackage[normalem]{ulem}
\usepackage{graphics}
\usepackage{tikz}
\usepackage{array}
\usepackage{animate}
\usepackage{titlesec}
\usepackage{balance}
\usepackage{enumitem}
\usepackage{booktabs}
\usepackage{comment}

\titlespacing*{\section}{0pt}{3pt}{3pt}
\titlespacing*{\subsection}{0pt}{3pt}{2pt}
\titlespacing*{\subsubsection}{0pt}{3pt}{2pt}

\copyrightyear{2024}
\acmYear{2024}
\setcopyright{acmlicensed}
\acmConference[DAC '24]{61st ACM/IEEE Design Automation Conference}{June 23--27, 2024}{San Francisco, CA, USA}
\acmBooktitle{61st ACM/IEEE Design Automation Conference (DAC '24), June 23--27, 2024, San Francisco, CA, USA}
\acmDOI{10.1145/3649329.3657343}
\acmISBN{979-8-4007-0601-1/24/06}

\begin{document}

\title{GreenFPGA: Evaluating FPGAs as Environmentally \\ Sustainable Computing Solutions}
\vspace{-1mm}
\author{Chetan Choppali Sudarshan, Aman Arora, and Vidya A. Chhabria}
\affiliation{Arizona State University}

\begin{abstract}
\noindent
Growing global concerns about climate change highlight the need for environmentally sustainable computing. The ecological impact of computing, including operational and embodied, is crucial. Field Programmable Gate Arrays (FPGAs) stand out as promising sustainable computing platforms due to their reconfigurability across various applications. This paper introduces GreenFPGA, a tool estimating the total carbon footprint (CFP) of FPGAs over their lifespan, considering design, manufacturing, reconfigurability, operation, disposal, and recycling. Using GreenFPGA, the paper evaluates scenarios where the ecological benefits of FPGA reconfigurability outweigh operational and embodied carbon costs, positioning FPGAs as an environmentally sustainable choice for hardware acceleration compared to Application-specific integrated circuits (ASICs). Experimental results show that FPGAs have lower CFP than ASICs for multiple low-volume applications or short application lifespans.

\end{abstract}	

\maketitle

\vspace{-3mm}
\section{Introduction} 
\label{sec:intro}
\noindent
From microchips to data centers, all computing has a carbon footprint (CFP) with significant environmental impact. The growing demand for computational power, driven by applications like artificial intelligence\cite{green-ml-1}, has led the information and computing technology (ICT) sector to contribute 2.1\% to 3.9\% of the world's total CFP~\cite{Freitag2021TheRC}. Traditionally, the semiconductor industry prioritized making chips smaller, faster, and more energy-efficient, focusing on reducing operational CFP. However, the ecological impact of design, manufacturing, and disposal, known as embodied carbon, is equally crucial. The importance of embodied carbon is evident from a surge of interest in sustainable computing from the US government \cite{nsf_dear_colleague_sustainability}.

Previous studies~\cite{act, sudarshan2023ecochip} emphasize the significant role of embodied carbon in the total emissions of modern data centers and edge devices. They offer analysis tools for calculating carbon footprint (CFP) from manufacturing at the computer architecture level, addressing both monolithic and chiplet-based systems. However, these approaches do not extend to field programmable gate arrays (FPGAs), reconfigurable computing platforms. While existing work embraces the 3R concept (reduce-reuse-recycle), as seen in~\cite{sudarshan2023ecochip}, showcasing the ``reuse" of predesigned chiplets for sustainable computing, this paper leverages a fourth ``R." This new focus is on reconfigurable devices (FPGAs) and their potential to reduce the overall CFP~\cite{fpga-sustainability}.

\begin{figure}[t]
\centering
\includegraphics[width=0.95\linewidth]{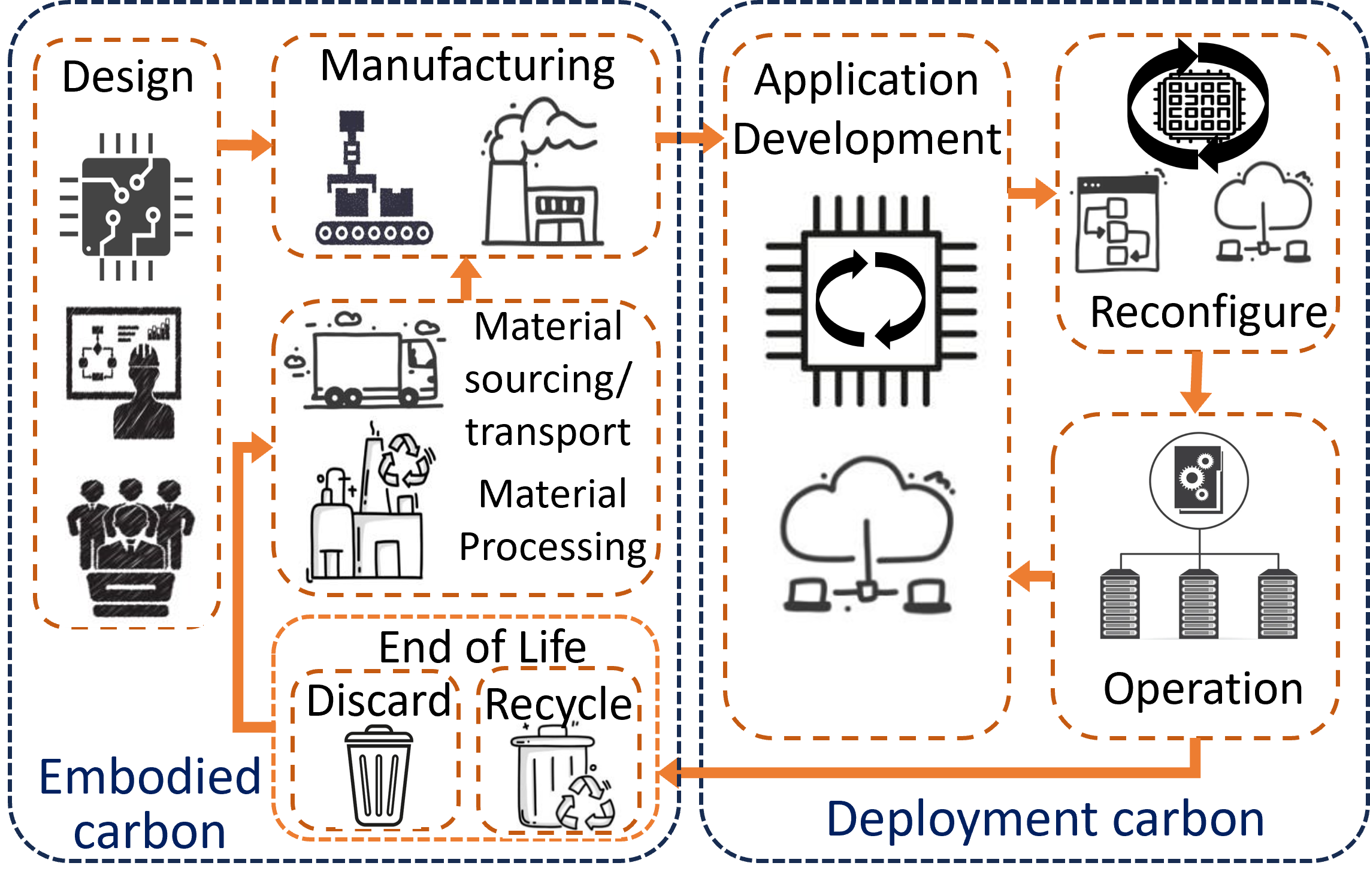}
\vspace{-6mm}
\caption{Lifecycle of an FPGA: Highlighting the embodied and operational CFP from cradle (design) to grave (disposal).}
\vspace{-6mm}
\label{fig:lifecycle}
\end{figure}

With Moore's Law slowing down, hardware acceleration for computationally intensive tasks is now widespread. Three key options for hardware acceleration are GPUs, FPGAs, and Application-Specific Integrated Circuits (ASICs). FPGAs, known for their reconfigurability and energy efficiency, excel at speeding up emerging and ever-changing workloads like machine learning\cite{Sevilla_Heim_Ho_Besiroglu_Hobbhahn_Villalobos_2022}. They are extensively used in both cloud~\cite{microsoft_reconfig_datacenter} and edge computing~\cite{Biookaghazadeh_Zhao_Ren}, forming a significant part of global computing infrastructure. In this paper, we focus on evaluating the CFP of FPGAs as they are potentially greener alternatives for hardware acceleration. We develop new models that can account for these FPGA-related distinctions. We perform comparisons against ASICs as accelerating alternatives as CPUs are inefficient for compute-intensive applications and GPUs have high-power and less flexibility than FPGAs.

Fig.~\ref{fig:lifecycle} shows the complete lifecycle of an FPGA, highlighting its ability to be reconfigured for different applications. The figure outlines the factors contributing to embodied carbon footprint (CFP), including design, manufacturing, discard, and recycling, while operational CFP is tied to end-user activities. The key difference compared to ASICs is that FPGAs can be reused across diverse applications, whereas ASICs reach the end of their lifecycle once the application's lifetime is complete. In this paper, we introduce GreenFPGA, a tool to assess the CFP of FPGA-based computing throughout its lifecycle. Using GreenFPGA, we analyze scenarios to determine when FPGAs are more sustainable computing and acceleration platforms than ASICs. The key contributions are:

\vspace{-1mm}
\begin{enumerate}
    \item To the best of our knowledge, GreenFPGA is the first to model and assess the CFP of FPGA across its entire lifespan.

    \item We develop a robust model for CFP of the design phase of a chip's lifecycle, compared to~\cite{sudarshan2023ecochip} based on industry reports.

    \item GreenFPGA accounts for the unique aspects of FPGA-based computing, including CFP overheads from reconfiguring the FPGA and application development time, unlike~\cite{sudarshan2023ecochip}. 
   
    \item We compare ASICs and FPGAs and list scenarios under which FPGAs are more sustainable. 
    
    \item GreenFPGA indicates FPGAs are sustainable alternatives to ASICs in three scenarios: (i) for application lifetimes below 1.6 years, (ii) when the FPGA is used in over five applications, or (iii) when application volumes are under 2 million for iso-performance in specific domains.
   
\end{enumerate}

\noindent
GreenFPGA is open-source and available to the public~\cite{github_greenfpga}. 
\section{FPGAs as a sustainable computing solution}
\label{sec:preliminaries}

\begin{figure}[t]
\centering
\includegraphics[width=0.7\linewidth]{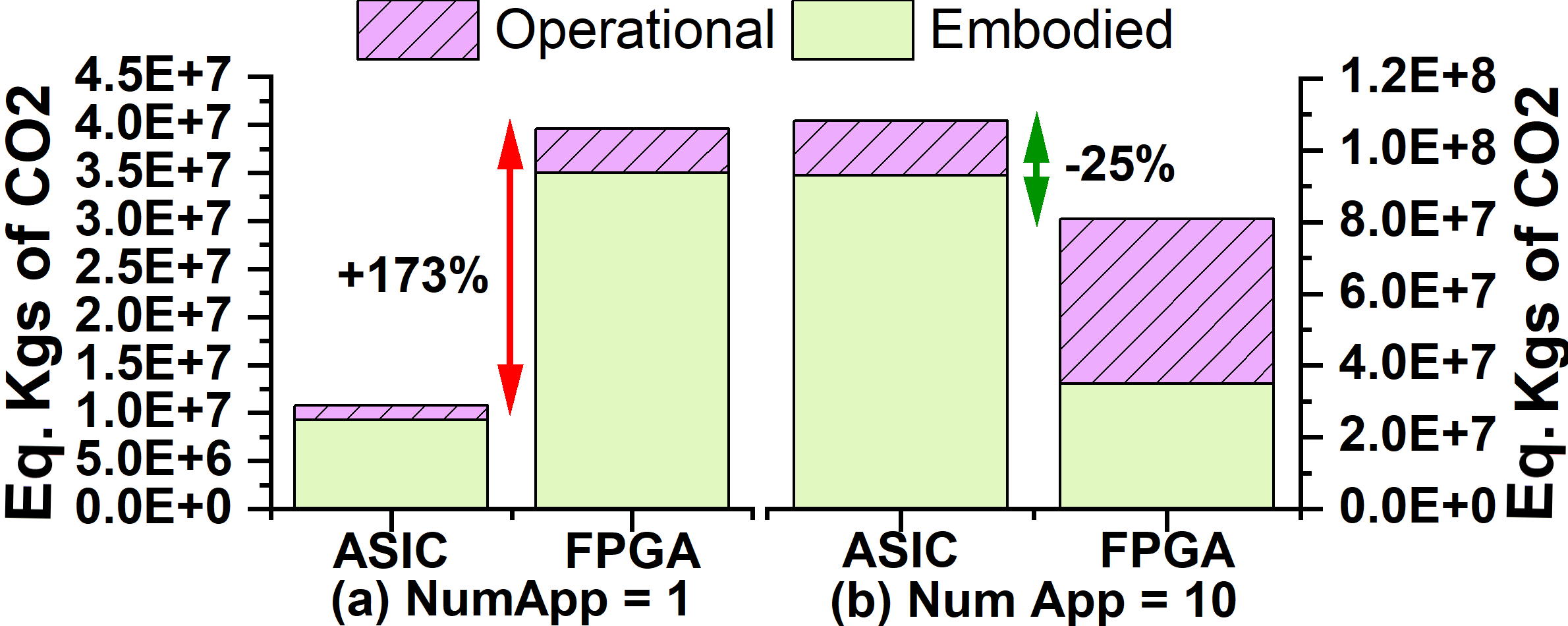}
\vspace{-4mm}
\caption{CFP comparison between ASIC and FPGA-based computing for a single application and ten applications.}
\vspace{-7mm}
\label{fig:motivation}
\end{figure}

\noindent
{\bf Scope for sustainable computing via FPGAs}
While FPGAs exhibit larger physical size and lower energy efficiency compared to equivalent-performing ASICs, leading to increased embodied and operational carbon footprints (CFP), they offer compelling environmental sustainability advantages:
\begin{itemize}
\item Reconfigurability: FPGAs stand out by being reconfigurable and adaptable for multiple applications post-manufacturing, a unique feature absent in ASICs.
\item  Extended lifespan: FPGAs typically have longer lifespans and use cases lasting 12 to 15 years~\cite{altera_fpga_lifetime} as they can be reconfigured, compared to ASIC that become obsolete with rapidly changing application workloads (5 to 8 years).
\end{itemize}
\noindent
Over time and across diverse applications, the embodied CFP incurred during FPGA manufacturing can be amortized across its extended lifespan and various uses, unlike ASICs. Fig.~\ref{fig:motivation} illustrates this point with an FPGA at iso-performance\cite{tian_tan_phd_utexas} with ASIC for deep neural network (DNN) domain. Although the FPGA initially has a higher CFP than the ASIC due to size and energy efficiency differences, when reused for ten different applications at iso-performance, the FPGA's recurring embodied CFP is saved, resulting in a 25\% lower CFP compared to the ASIC. In the context of rapidly changing and compute-intensive workloads, such as machine learning, FPGAs emerge as sustainable alternatives for hardware acceleration.


\noindent
{\bf Existing models for CFP estimation} Three prior bodies of work focus on CFP estimation at the architectural level for ASICs and CPUs: the first body of work includes~\cite{firstorder-simple, kaya-ghent}, the second includes~\cite {chasing-carbon, act, act+}, and the third includes~\cite{sudarshan2023ecochip,3dcarbon}. The work in~\cite{kaya-ghent} reformulated the Kaya identity to understand how the global CFP of computer systems evolves and has made a case to lower chip sizes to lower embodied CFP and~\cite{firstorder-simple} creates a simple model based on first principles. The works in~\cite{act, chasing-carbon, act+} have created data-driven model, from publicly available sustainability reports from industry~\cite{apple-report, TSMC-report, imec-wp, amd-report}, for embodied carbon estimation and have created a platform for carbon-aware design space exploration (DSE)~\cite{act+}. The works in~\cite{sudarshan2023ecochip, 3dcarbon}, build on~\cite{act} and~\cite{imec-dtco}, to measure the CFP of the 2.5D/3D heterogeneously integrated VLSI systems. The total CFP of ASICs to perform $N_\text{app}$ different applications is~\cite{sudarshan2023ecochip}:

\vspace{-2mm}
\begin{equation}
     C_\text{ASIC} = \sum^{i=N_\text{app}}_{i=1}  (C_\text{emb, i} +  T_i \times C_\text{deploy, i}) 
    \label{eq:tot-cfp-asic}
\vspace{-1mm}
\end{equation}

\noindent
where the embodied CFP, $C_\text{emb, i}$ is given by the sum of the CFP from design, manufacturing, and packaging, for application $i$, 
$T_\text{i}$ is the lifetime of the application $i$ 
and $C_\text{deploy, i}$ is the CFP from application development and operation of the ASIC for application $i$ in the field. The models for design, manufacturing, and packaging are available in~\cite{sudarshan2023ecochip}. However, these models do not directly apply to FPGAs which are particularly interesting to model for CFP due to their reconfigurability. In GreenFPGA, we improve the models for the design CFP and also incorporate new models for end-of-life (discard and recycling), and software-related ASIC programming for the ASIC, and hardware reconfiguration for the FPGA.

\section{ GreenFPGA: Models for CFP assessment}
\begin{figure}[t]
\centering
\includegraphics[width=\linewidth]{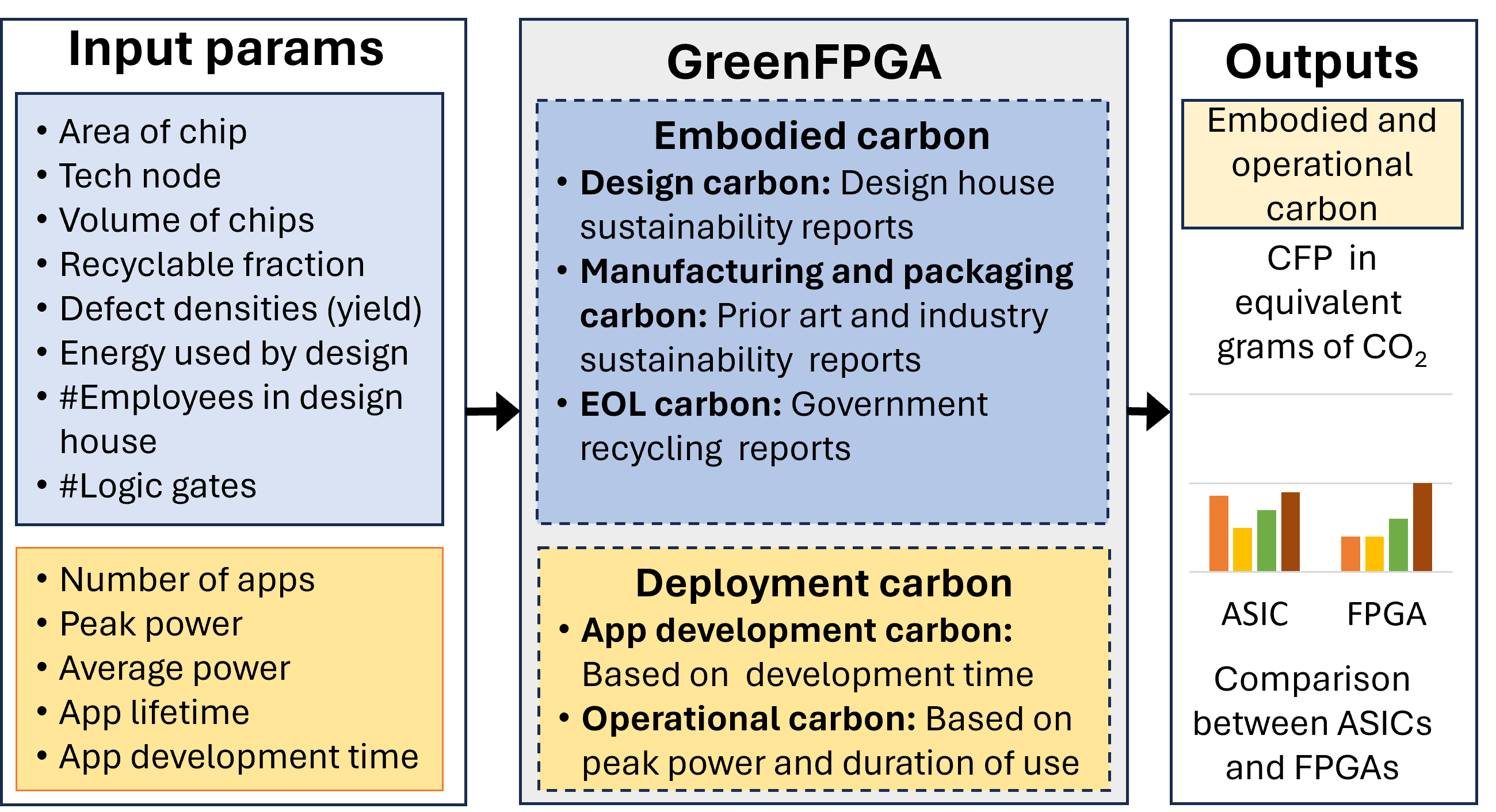}
\vspace{-6mm}
\caption{GreenFPGA: inputs, outputs, and models.}
\vspace{-7mm}
\label{fig:top-level}
\end{figure}

A high-level description of GreenFPGA is shown in Fig.~\ref{fig:top-level}, which highlights inputs, outputs, and the models used for the different components of CFP across the lifecycle. GreenFPGA presents a novel design CFP model and EOL CFP model from industry and government reports. The manufacturing and packaging model is utilized from prior art~\cite{act, imec-dtco,sudarshan2023ecochip}, which is based on industry reports. Further, we develop a model that accounts for reconfigurability and application development time specific to FPGAs. The inputs highlighted in blue are specific to embodied CFP, and the ones in yellow are for deployment carbon estimation. GreenFPGA generates the CFP of FPGA and ASIC-based computing and shows scenarios where FPGAs are more sustainable than  ASICs at iso-performance.

\subsection{GreenFPGA: Total CFP model} 
Given that both embodied CFP and operational CFP contribute to the total CFP, we model the total CFP of the FPGA as the sum of the embodied and operational CFP as shown below. However, unlike ASICs (Equation~\eqref{eq:tot-cfp-asic}), the same FPGA can be reused for different applications, and therefore, the total CFP for $N_\text{app}$ applications using FPGAs is given by  $C_\text{FPGA}$: 

\vspace{-3mm}
\begin{equation}
     C_\text{FPGA} =   C_\text{emb} + \sum^{i=N_\text{app}}_{i=1}  T_i \times  C_\text{deploy, i} 
    \label{eq:tot-cfp-fpga}
\end{equation}

\noindent 
where $C_\text{emb}$ is the embodied CFP of FPGA, 
$T_i$ is the  lifetime of application $i$, and
$C_\text{deploy, i}$ is the deployment CFP of the FPGA for application $i$. 
In the rest of this section, we detail the models for each of the components in Eq.~\eqref{eq:tot-cfp-fpga}.


\subsection{GreenFPGA: Embodied CFP model}
With embodied CFP dominating the total CFP, particularly in battery-operated devices, and devices on the edge~\cite{chasing-carbon}, it is crucial to have models that account for CFP from different sources. The embodied CFP accounts for activities related to manufacturing, end-of-life, design, and packaging. The embodied CFP for the ASIC or FPGA, $C_\text{emb}$, for a application volume of $N_\text{vol}$ chips is given by: 

\vspace{-3mm}
\begin{equation}
\label{eq:embodied}
    C_\text{emb}= C_{\text{des}} +  N_\text{vol} \times N_\text{FPGA} \times (C_{\text{mfg}} + C_\text{package} + C_\text{EOL})
\end{equation}

\noindent 
where $C_{\text{des}}$ is the design CFP that accounts for activities related to chip design and test, $C_\text{mfg}$ is the manufacturing CFP that accounts for activities related to the fab,  and $C_{\text{package}}$ is the CFP from package manufacture and assembly of the FPGA or ASIC, and $C_\text{EOL}$ is the EOL CFP to model recycle and discard activities. For certain applications, iso-performance comparisons between ASICs require more than one FPGA, as the ASIC counterparts are either at reticle limits or have extremely high performance. Therefore,  we define $N_\text{FPGA}$ as the number of FPGAs required for a given application for iso-performance to the ASIC and is given by $  \left\lceil\frac{\text{app}_\text{size}}{\text{FPGA}_\text{capacity}}  \right\rceil$ where the application size and FPGA capacity are specified in terms of equivalent logic gates\footnote{For ASIC $C_\text{emb}$, $N_\text{FPGA} = 1$ such that we can reuse the same model.}.

\noindent
{\em (1) Design CFP:} The activities related to chip design (ASIC or FPGA) include architectural development, RTL, verification, synthesis, simulations, place and route, and various analyses, test and post-silicon validation. These activities are performed by large design houses where several engineers work on a common product. While the only existing prior art has used a simplified model that relied on the number of logic gates~\cite{sudarshan2023ecochip} only, the model is difficult to validate. GreenFPGA models the design CFP based on the energy usage of large design houses obtained sustainability reports~\cite{amd-report, microchip-report, nvidia-report}, the number of products (chips) designed, the fraction of energy coming from renewable resources, the number of employees working on the specific product, and the size of the chip (number of logic gates). The design CFP $C_\text{des}$ for an FPGA or ASIC is given by: 

\vspace{-3mm}
\begin{equation}
    C_\text{des} = C_\text{emp} \times N_\text{emp, des}  \times \frac {N_\text{gates}}{N_\text{gates, des}}  \times T_\text{proj}
    \label{eq:des}
\end{equation}

\noindent
where $C_\text{emp}$ is the CFP per employee per year, the $N_\text{emp, des}$ is the average number of employees per chip to be designed,  and $N_\text{gates}$ is the number of logic gates in the chip, and $N_\text{gates,  prod}$ is the average number of gates per chip, and $T_\text{proj}$ is the duration the chip design project. The $C_\text{employee}$ is obtained from industry reports of fabless design houses and is given by $C_\text{employee} = E_\text{des} \times C_\text{src, des}$ where $E_\text{des}$ is the electric energy utilized by the design house per year and $C_\text{src, des}$ is the carbon intensity for the energy source. $C_\text{src, des}$ will be lower for renewable resources and higher for non-renewable resources~\cite{chasing-carbon}. The energy utilized by the project over its duration will include all design-related activities and testing. The model in~\cite{sudarshan2023ecochip}, did not account for testing, validation and was difficult to validate the true contribution to design CFP. Similar to manufacturing CFP, GreenFPGA also models the design CFP from industry sustainability reports~\cite{nvidia-report, microchip-report, amd-latest-report}, which is a more reliable source as it utilizes energy values mentioned in the reports.

\noindent
{\em (2) Manufacturing CFP incorporating recycled materials}:
GreenFPGA employs manufacturing CFP models from~\cite{sudarshan2023ecochip, imec-dtco, act}, where $C_\text{mfg}$ includes manufacturing from sourcing materials ($C_\text{materials}$), greenhouse gas emissions, and energy utilization by the fab. These models utilize data from industry reports~\cite{apple-report,facebook-report} and papers~\cite{imec-wp, imec-dtco}. GreenFPGA models the materials fetched from recycled sourcing and newly extracted materials. We utilized data from~\cite{apple-recycle,recycle-cfp} to extract the percentage of materials that can be recycled ($\rho$) from products and scale the CFP from sourcing materials as:

\vspace{-5mm}
\begin{equation}
\label{eq:mfg-tot}
    C_\text{materials} = {\rho} C_\text{materials, recycled} + (1-{\rho})  C_\text{materials, new}
\end{equation}

\noindent
where $C_\text{materials}$ is the component of $C_\text{mfg}$ related to sourcing of raw materials, and $C_\text{materials, new}$ is the manufacturing CFP when the materials are extracted from the source, and $C_\text{materials, recycled}$ is the CFP from utilizing recycled materials. 
The rest of the components as a part of $C_\text{mfg}$ are modeled in the same way as~\cite{sudarshan2023ecochip}.

\noindent
{\em(3) Package CFP:} We use the monolithic package CFP model from~\cite{sudarshan2023ecochip}. 

\noindent
{\em(4) EOL CFP:} The end-of-life CFP includes CFP from discarding and a CFP credit for recycling a fraction ($\delta$) of the chip and is given by:

\vspace{-3mm}
\begin{equation}
\label{eq:eol}
C_\text{EOL} =  (1-\delta ) C_\text{dis} - \delta C_\text{recycle} 
\end{equation}

\noindent
where $\delta$ $C_\text{recycle}$ is the CFP recycling credit, and $C_\text{dis}$ is the CFP of discarding. These values are obtained from government reports~\cite{epa-gov-recycle}.

\subsection{GreenFPGA: Deployment CFP model}
GreenFPGA models the deployment CFP, $C_\text{deploy}$, as the sum of the CFP from field operation (product use) and the application development and is given by: $C_\text{deploy}= N_\text{vol} \times C_\text{op}  + C_\text{app-dev}$
where  $C_\text{op}$ is the operational CFP during use of the chip, $C_\text{app-dev}$ is the application development CFP, and $N_\text{vol}$ is the number of chips (ASICs or FPGAs manufactured).




\noindent
{\em(1) Operational CFP:} The operational CFP, $C_\text{op}$, is modeled as the product of carbon intensity of the source of energy during usage ($C_\text{src, use}$) and the energy spent during usage ($E_\text{use}$) and is given by $ C_\text{op}= C_\text{src, use}  \times E_\text{use} $ where the energy spent during usage is a function of peak power and duty cycles~\cite{sudarshan2023ecochip}. 

\noindent
{\em(2) Application development CFP:}
For FPGAs, application development involves RTL (Register Transfer Level) development or HLS (High level synthesis) flows and hardware-level simulations. In contrast, ASICs utilize software flows with extensive regression testing, as seen in the Google TPU \cite{tpuv4}, if at all. These different approaches lead to distinct CFPs, because software development is faster than hardware development. We consider this overhead when assessing FPGAs as sustainable computing solutions, as application development represents a recurring cost per application in Eq.~\eqref{eq:tot-cfp-fpga}. We model $C_\text{app-dev}$ as the product of the power dissipated by the CPU systems used in application development, the carbon intensity of the energy source, and the application development time, $T_\text{app-dev}$:

\vspace{-5mm}
\begin{equation}
\label{eq:appdev-cfp}
    T_\text{app-dev} = N_\text{app} \times ( T_\text{app, FE}  + T_\text{app, BE} ) + N_\text{vol} \times T_\text{app,config}  
\end{equation}

\noindent
where $N_\text{app}$ is the total number of applications, $T_\text{app,FE}$ is the time it takes to write RTL and perform verification which is done once per application, $T_\text{app,BE}$ is the time it takes to synthesize, place and route which is usually performed once per FPGA architecture, $N_\text{vol}$ is the volume of FPGAs manufactured and $T_\text{app-config}$ is time it takes to configure the FPGA that is deployed in the field.  For ASICs, $T_\text{app, FE}$ and $T_\text{app, BE}$ are zero, as they are already accounted for in Eq.~\eqref{eq:des}.

\section{Evaluation of GreenFPGA}
\label{sec:results}

\begin{table}[tb]
\centering

\caption{Input parameters ranges to GreenFPGA}
\label{tbl:params}
\vspace{-3mm}
\resizebox{0.9\linewidth}{!}{%
\begin{tabular}{c|crll}
\hline
Model   &     Parameter     &     Value&Unit     &     Source   \\ \hline

 
 \multirow{1}{*}{$C_\text{materials}$}    
 &  $\rho$      &     0 -- 1  &   &  \cite{apple-recycle}/user-defined    \\
 \hline

 \multirow{3}{*}{$C_\text{EOL}$}    
 &  $\delta$      &     0 -- 1  &   &  \cite{epa-gov-recycle}   \\
 &  $C_\text{recycle}$     &    7.65 -- 29.83 & $\text{MTCO}_\text{2}\text{E/ton}$ & \cite{epa-gov-recycle}    \\
 &  $C_\text{dis}$         &    0.03 -- 2.08  & $\text{MTCO}_\text{2}\text{E/ton}$ & \cite{epa-gov-recycle}    \\
 \hline
 
 \multirow{3}{*}{$C_\text{app-dev}$}    
 & $T_\text{app, FE}$     &    1.5 -- 2.5 & months &  user-defined  \\
 &  $T_\text{app, BE}$     &    0.5 -- 1.5 & months &   user-defined \\
 \hline

 \multirow{3}{*}{$C_\text{des}$}    
 &  $E_\text{des}$          &    2 --  7.3            &  GWh &  \cite{amd-latest-report,nvidia-report,microchip-report}   \\
 & $C_\text{src, des}$ & 30-700 &  g CO$_2$/kWh   &  \cite{act,imec-dtco} \\
 &  $N_\text{emp, des}$     &    20K -- 160K & employees &   \cite{microchip-report,nvidia-report,amd-latest-report} \\
 &  $T_\text{proj}$         &    1 -- 3    & years &  \cite{nvidia-roadmap}  \\
 \hline
 
\end{tabular}%
}
\vspace{-3mm}
\end{table}


\subsection{Experimental setup and testcases}
\noindent
{\bf Input parameters} GreenFPGA uses several input parameters described in the paper and listed in Table~\ref{tbl:params} with their corresponding sources. Certain parameters, such as the application development time, are difficult to find in the public domain. We assume values for these based on industry experience. However, the user can tune these. Further, GreenFPGA also relies on other parameters from~\cite{sudarshan2023ecochip, act}, which are utilized as-is from their GitHub repositories~\cite{ecochip-github, act-github} for manufacturing and packaging CFP.

\noindent
{\bf Testcases}  To assess FPGAs as sustainable alternatives to ASICs, we compare them at iso-performance, employing power, and area values for across three domains (Deep Neural Networks (DNN), Image Processing (ImgProc), Cryptography (Crypto)) from~\cite{tian_tan_phd_utexas} considering testcases in a 10nm technology node.~\cite{tian_tan_phd_utexas} emphasizes that while ASICs are designed for flexibility and programmability, they lack reconfigurability at the circuit level post-manufacturing. In contrast, FPGAs offer circuit-level reconfigurability, allowing the architecture of an application to be finely tuned to specific requirements. This insight results in practical ratios of area and power metrics between FPGAs and ASICs for the same performance, as highlighted in Table~\ref{tbl:experiment-testcases}. The assumption is that FPGAs can adapt to changing applications by loading new configurations, whereas a new ASIC is required for each application change.

Further, we evaluate GreenFPGA on four industry testcases as listed in Table~\ref{tbl:industry-testcases} including two ASICs IndustryASIC1 and IndustryASIC2 based on Moffett Antoum~\cite{moffett} and Google TPU~\cite{tpuv4}, respectively, and two FPGAs, IndustryFPGA1 and IndustryFPGA2 based on Intel Agilex 7~\cite{agilex7} and Stratix 10~\cite{stratix10}, respectively. The table lists the values of power (TDP), area, and technology nodes.

\begin{table}[t]
  \centering
  \caption{FPGA testcases for iso-performance as  ASIC~\cite{tian_tan_phd_utexas}}
\vspace{-4mm}
  \resizebox{0.8\linewidth}{!}{
  \begin{tabular}{|c|c|c|c|}
    \hline
    \textbf{Testcases} & \textbf{DNN} & \textbf{ImgProc} & \textbf{Crypto}  \\
    \hline
    Area (normalized to ASIC)      & 4 & 7.42 & 1 \\
    \hline
    Power (normalized to ASIC)     & 3 & 1.25 & 1 \\
    \hline
  \end{tabular}
  }
  \label{tbl:experiment-testcases}
   \vspace{-4mm}
\end{table}

\begin{table}[t]
  \centering
  \caption{Summary of industry testcases~\cite{moffett, stratix10, agilex7, tpuv4}}
  \vspace{-4mm}
  \resizebox{\linewidth}{!}{%
  \begin{tabular}{|c|c|c|c|c|}
    \hline
    \textbf{Testcases} & \textbf{IndustryASIC1} & \textbf{IndustryASIC2} & \textbf{IndustryFPGA1} & \textbf{IndustryFPGA2} \\
    \hline
    Area      & 340 $mm^2$ & 600 $mm^2$ & 380 $mm^2$ & 550 $mm^2$ \\
    \hline
    Power     & 70 W 	& 192 W  & 160 W  & 220 W \\
    \hline
    Tech. Node & 12 nm 	& 7 nm   & 14 nm  & 10 nm \\
    \hline
  \end{tabular}
  }
  \label{tbl:industry-testcases}
    \vspace{-4mm}
\end{table}

\subsection {Comparing CFP of FPGAs and ASICs}

As noted earlier, the embodied CFP and the deployment CFP of an FPGA is higher than that of an iso-performance ASIC, because the FPGA required has a larger area and consumes more power.
However, FPGA reconfigurability can help amortize the embodied CFP over the chip's lifetime, thereby reducing the overall CFP.
In this section, we observe the impact of number of applications (Num Apps) $N_\text{app}$, volume of applications (App Volume)  $N_\text{vol}$, and lifetime of each application (App Lifetime) $T_\text{i}$, on the CFP of FPGAs and ASICs.
For these experiments, iso-performance ASICs and FPGAs are used~\cite{tian_tan_phd_utexas}.
For each experiment, we define the point at which the CFP of FPGA becomes lower compared to ASIC as the \textit{A2F crossover point}, and the point at which the CFP of FPGA becomes higher compared to the ASIC as the \textit{F2A crossover point}.

\begin{figure}[t]
\centering
\includegraphics[width=\linewidth]{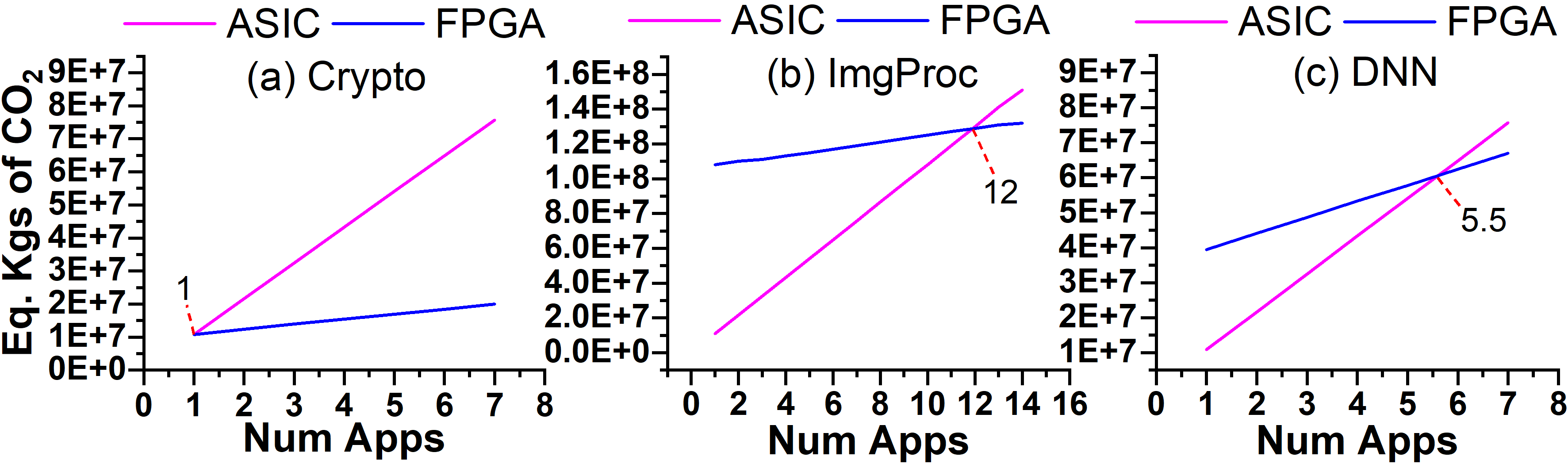}
\vspace{-8mm}
\caption{Variation of CFP with $N_\text{app}$; $N_\text{vol}$ and $T_\text{i}$ are constant.}
\vspace{-4mm}
\label{fig:cfp-vs-napp}
\end{figure}

\begin{figure}[t]
\centering
\includegraphics[width=\linewidth]{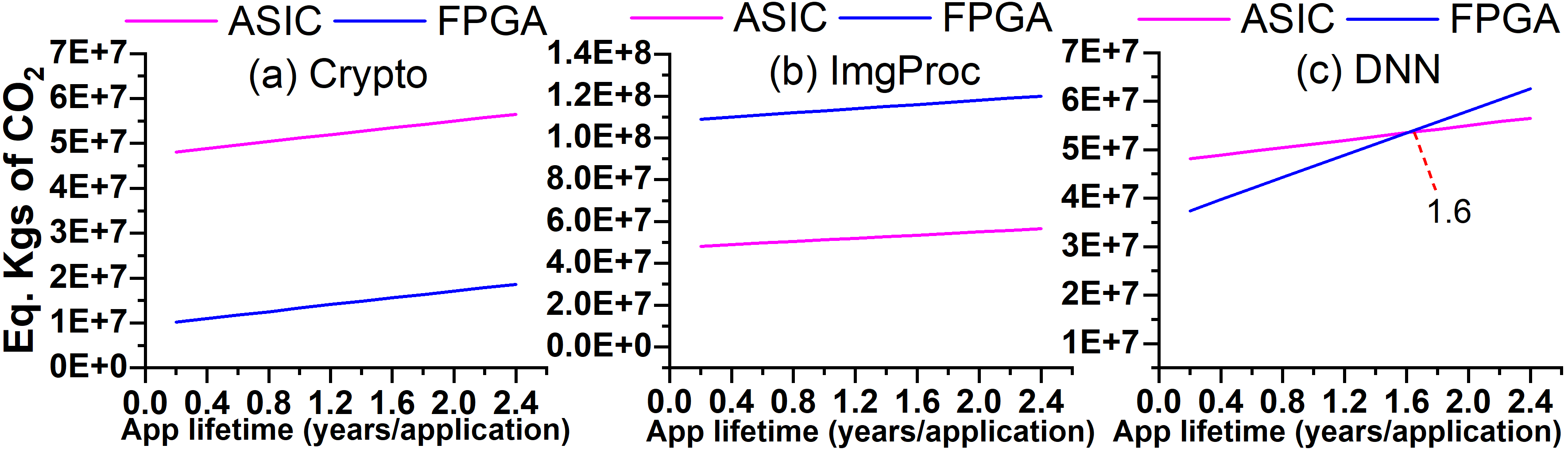}
\vspace{-8mm}
\caption{Variation of CFP with $T_\text{i}$; $N_\text{vol}$ and $N_\text{app}$ are constant.}
\vspace{-4mm}
\label{fig:cfp-vs-app-lifetime}
\end{figure}

\begin{figure}[t]
\centering
\includegraphics[width=\linewidth]{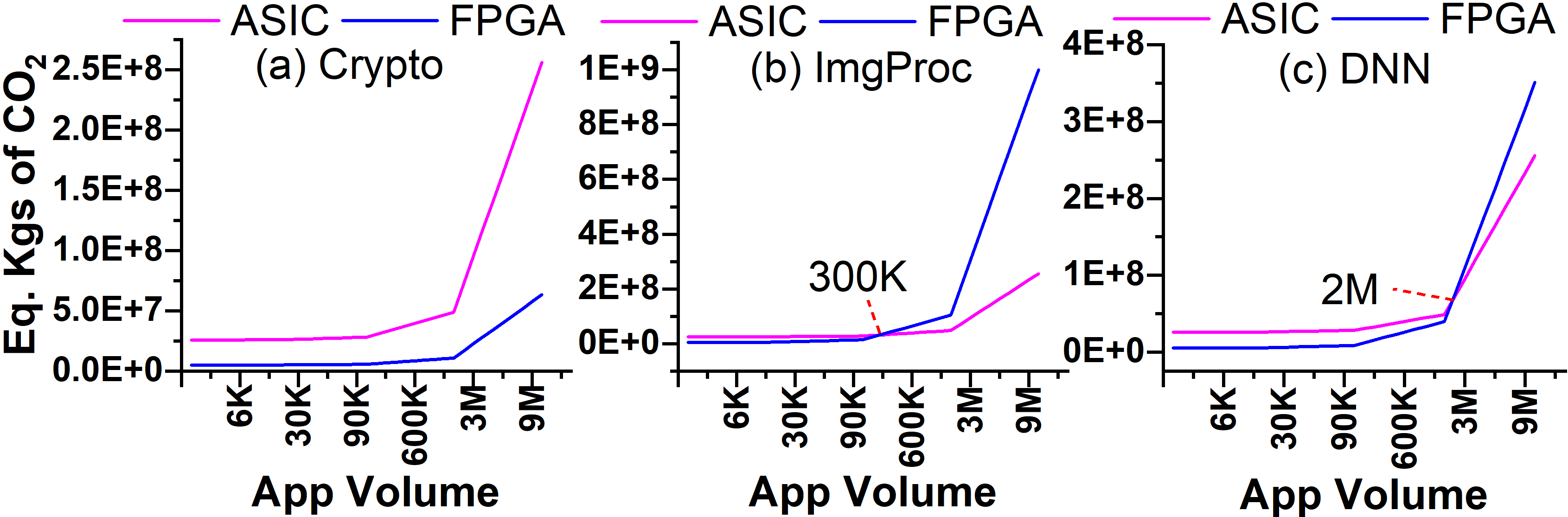}
\vspace{-8mm}
\caption{Variation of CFP with $N_\text{vol}$; $N_\text{app}$ and $T_\text{i}$ are constant.}
\vspace{-6mm}
\label{fig:cfp-vs-nvol}
\end{figure}

\noindent\textbf{(A) Impact of number of applications}:
We set up an experiment where the number of applications, $N_\text{app}$, is varied from 1 to 8, and
application lifetime, $T_i$ = 2 years and application volume $N_\text{vol}$ = 1e6 units. 
After the lifetime of an application, new ASIC chips need to be manufactured to support the new application, but FPGA chips can be reconfigured and redeployed.
The results of this experiment are shown in Fig.~\ref{fig:cfp-vs-napp}.
Notably, different application domains show different behavior, because iso-performance FPGA for each domain has a different area and power consumption (Table~\ref{tbl:experiment-testcases}).
For Crypto applications, we observe that A2F crossover point is achieved after the first application, because the area and power of the FPGA and ASIC implementations are similar.
For ImgProc, the A2F crossover does not happen until $N_\text{app}$=8. So, we extend $N_\text{app}$ beyond 8, and observe that 12 applications are required in this case.
For DNN, the A2F crossover happens after 6 applications, i.e. after 12 years.

\noindent\textbf{(B) Impact of application lifetime}:
We set up an experiment where the application lifetime, $T_i$, is varied from 0.2 to 2.5 years, and
number of applications, $N_\text{app}$ = 5 and application volume $N_\text{vol}$ = 1e6 units. 
The results of this experiment are shown in Figure \ref{fig:cfp-vs-app-lifetime}.
Very different results are seen for the three application domains.
For Crypto, FPGAs are always more sustainable irrespective of the application lifetime.
For ImgProc, ASICs are always more sustainable irrespective of the application lifetime due to the large power and area overheads of the FPGA.
However, for DNNs, we observe that if the application lifetime is short, FPGA CFP is lower than ASIC, with an F2A point at about 1.6 years.

\noindent\textbf{(C) Impact of application volume}: 
For this experiment, we set the number of applications, $N_\text{app}$, to 5, and lifetime of each application, $T_i$ to 2 years, and
vary the volume of each application $N_\text{vol}$ from 1e3(1K) to 1e6(1M) instances.
The results of this experiment are shown in Fig.~\ref{fig:cfp-vs-nvol}.
As the volume increases, the CFP increases as expected.
For Crypto, FPGAs always remain the sustainable option with the ASIC CFP being higher even at lower volumes, because the iso-performance FPGA for Crypto have similar area and power values as the ASIC (Table~\ref{tbl:experiment-testcases}), and with the ability to reuse FPGA chips across applications, FPGA CFP is lower.
For ImgProc and DNN, an F2A crossover is observed at 300K and 2M instances indicating that FPGAs are sustainable for lower application volumes. 

\begin{figure}[t]
\centering
\includegraphics[width=\linewidth]{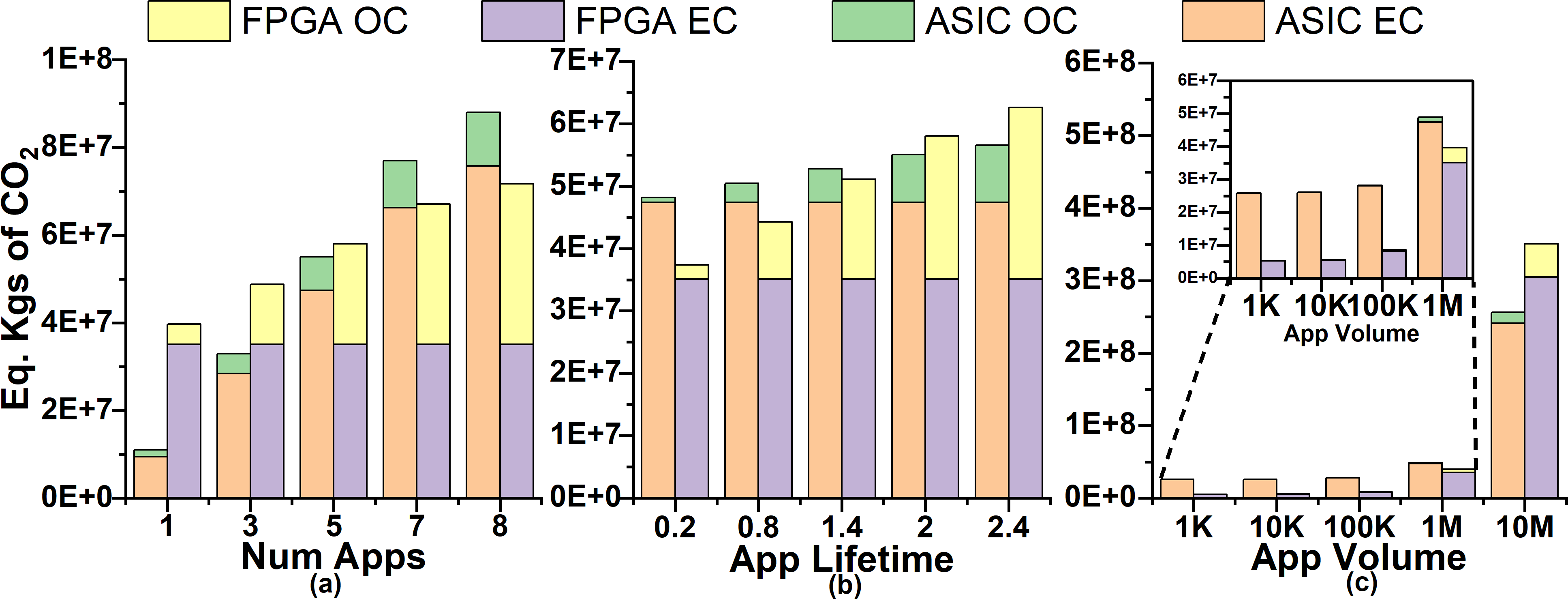}
\vspace{-7mm}
\caption{Different CFP components for the DNN domain with varying (a) $N_\text{app}$, (b) $T_i$, and (c) $N_\text{vol}$.}
\vspace{-6mm}
\label{fig:dnn-only}
\end{figure}

\noindent\textbf{(D) Deep dive into DNN domain's results}: 
In Fig.~\ref{fig:dnn-only}, we show the detailed breakdown of CFP of the three experiments listed above (A -- C) for the DNN application domain. 
The parameters used are $N_\text{app}$=5, $T_i$=2 years, $N_\text{vol}$=1e6 instances, unless that parameter is being varied. 
We analyze which components dominate the CFP - embodied CFP (EC)  or operational CFP (OC).
When $N_\text{app}$ is varied (Fig. \ref{fig:dnn-only}(a)), the EC of FPGAs stays the same, but OC increases as the number of applications increase. For ASICs, since new ASICs need to be manufactured for each application, EC increases significantly and dominates total CFP.
When $T_i$ is varied (Fig.~\ref{fig:dnn-only}(b)), the EC and OC stay the same, but with increased app lifetime, the FPGA OC begins to dominate, but the ASIC OC only increases marginally, making ASICs the better choice for longer app lifetimes.
When $N_\text{vol}$ is varied (Fig.~\ref{fig:dnn-only}(c)), for low volumes, EC dominates significantly and masking the OC component. The EC for ASICs is much higher than that of FPGAs because ASICs can not be re-configured for multiple applications. FPGA EC increases with increasing volume, and for large volumes, FPGAs are less sustainable than ASICs.

\begin{figure}[t]
\centering
\includegraphics[width=\linewidth]{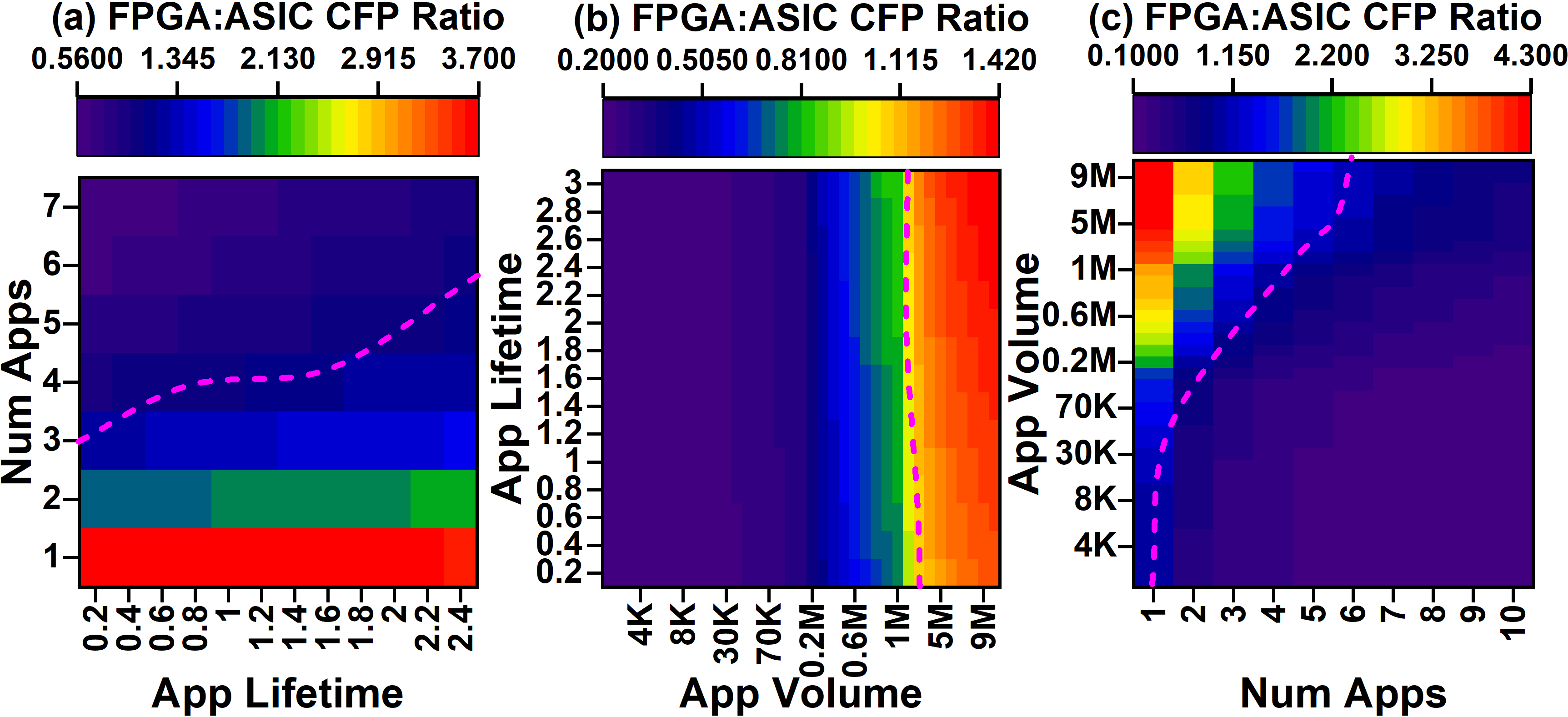}
\vspace{-8mm}
\caption{Variation in CFP for the DNN domain with pairwise sweep with (a) $N_\text{vol}$, (b) $N_\text{app}$, and (c)$T_i$ as constants.}
\vspace{-3mm}
\label{fig:heatmap-sweeps}
\end{figure}

Furthermore, to get more insight into the relationships of the three variables - $N_\text{vol}$, $N_\text{app}$ and $T_\text{i}$, we perform pairwise sweeps and generate heatmaps.
The results are shown in Fig.~\ref{fig:heatmap-sweeps}.
Each point on the heatmap shows the FPGA:ASIC CFP ratio.
These heatmaps help us understand the regions where ASICs are the more sustainable option (towards red) and where FPGAs are more sustainable (towards purple).
The crossover points are marked using pink dashes (FPGA:ASIC CFP ratio = 1).
Notably, for high app volumes (\verb|~|9M), FPGAs can be sustainable if number of applications is $>6$.
However, if the volume is high ($>3M$) or the number of applications is low ($<3$), then even lower lifetimes do not make FPGAs sustainable.

\begin{figure}[t]
\centering
\includegraphics[width=\linewidth]{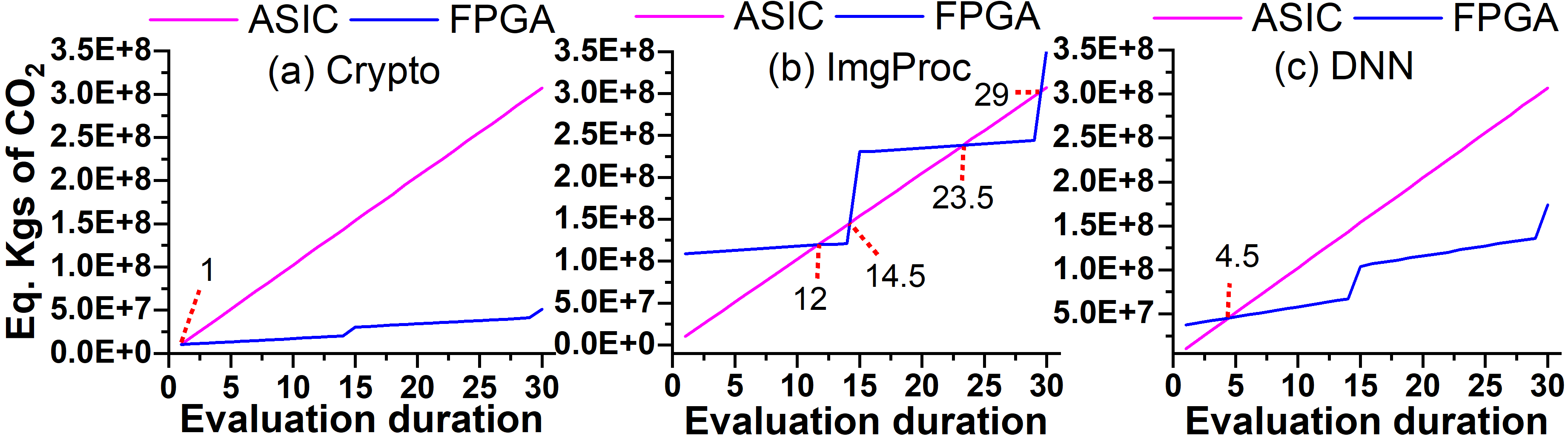}
\vspace{-6mm}
\caption{Variation in CFP with FPGA lifetime of 15 years and an application lifetime of 1 year.}
\label{fig:go-past-chip-life}
\vspace{-4mm}
\end{figure}

\noindent\textbf{(E) Increasing evaluation duration beyond chip lifetime}:
The evaluation period for experiment A was 15 years. However, for ImgProc (Fig.~\ref{fig:cfp-vs-napp}(b)), we went past 15 to find the crossover point.
However, if the chip's lifetime is 15 years, new chips must be manufactured every 15 years. This implies additional embodied carbon emissions.
To study the impact of this, we extend the duration of the experiment past the chip lifetime. The results are shown in Fig.~\ref{fig:go-past-chip-life}.
For all three application domains, noticeable increases (jumps) in the overall CFP can be seen at 15-year and 30-year marks, in the FPGA case.
On the contrary, no increases are seen in the ASIC because new chips must be manufactured based on application lifetime, rather than chip lifetime. For the ImgProc, these jumps lead to multiple A2F and F2A crossover points as the number of years of operation is increased, but for other domains, the choice for the more sustainable platform does not change.

\subsection{CFP estimation on industry testcases}

\begin{figure}[t]
\centering
\includegraphics[width=0.95\linewidth]{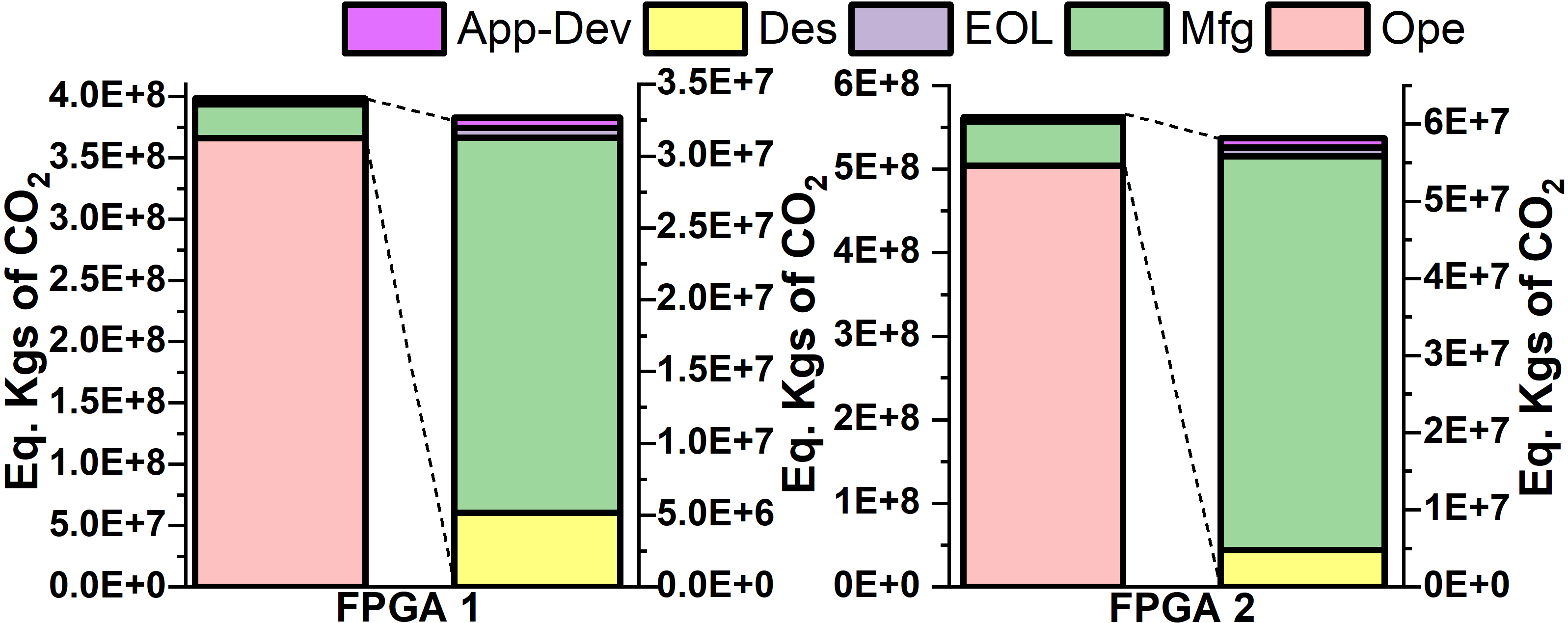}
\vspace{-4mm}
\caption{CFP for IndustryFPGA1 and IndustryFPGA2.}
\vspace{-7mm}
\label{fig:industry-fpga-cfp}
\end{figure}

{\bf Industry FPGAs:}  Fig.~\ref{fig:industry-fpga-cfp} highlights the CFP components of two industry FPGAs (Table~\ref{tbl:industry-testcases}) when each FPGA runs for six years with three applications and is reprogrammed thrice with a 1M volume using GreenFPGA. The CFP from application development is minimal even after reprogramming the FPGA three times, and it does not substantially contribute to the CFP overhead for both test cases. The primary contributor to the total CFP is the operational CFP, followed by the manufacturing and design CFP. Unlike prior art~\cite{sudarshan2023ecochip}, which grossly underestimated the design CFP, we utilize industry report energy value and find that design CFP to be 15\% of the embodied CFP. EOL CFP is a very small contributor.

\begin{figure}[t]
\centering
\includegraphics[width=0.95\linewidth]{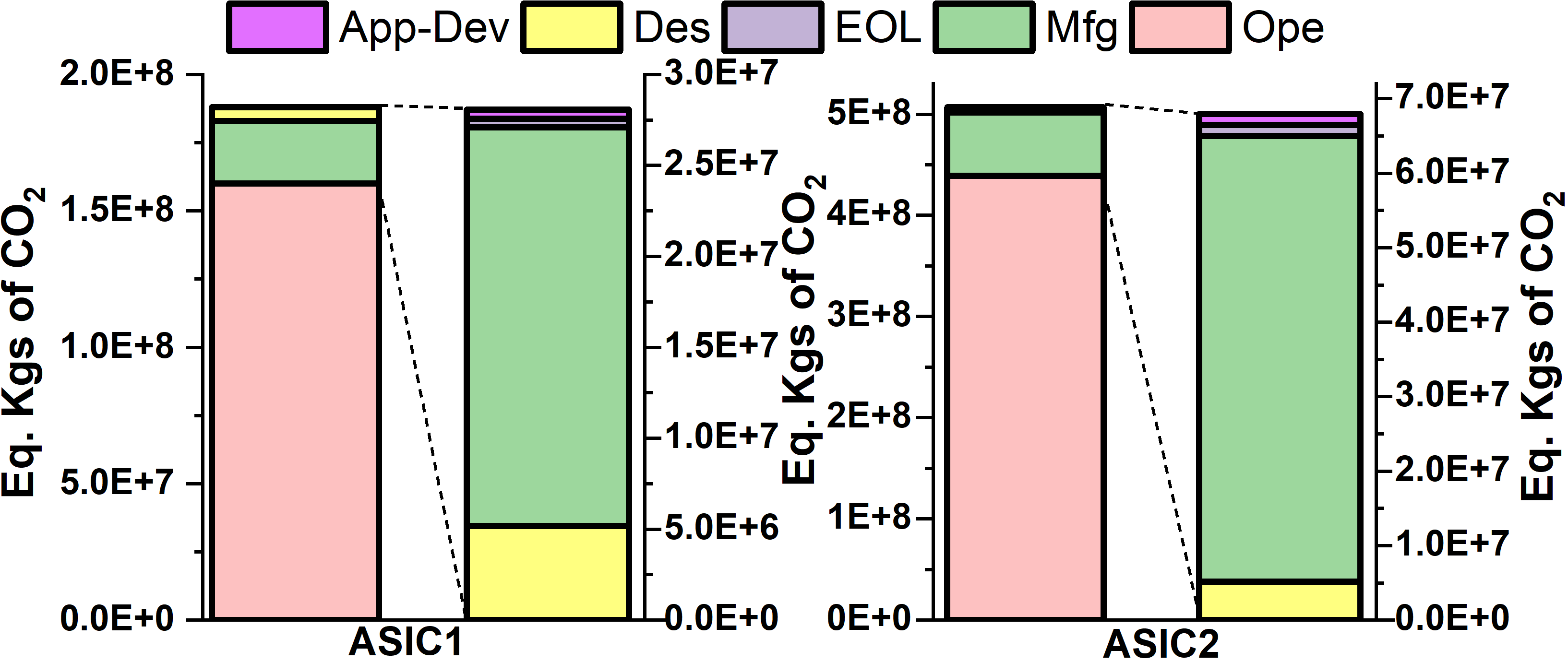}
\vspace{-4mm}
\caption{CFP for IndustryASIC1 and IndustryASIC2.}
\vspace{-5mm}
\label{fig:industry-asic-cfp}
\end{figure}

\noindent
{\bf Industry ASICs:} Fig.~\ref{fig:industry-asic-cfp} shows the different components of CFP evaluated using GreenFPGA on the two industry ASICs that were described in Table~\ref{tbl:industry-testcases}. The application time spans six years with a 1M volume, and in this scenario, the ASICs are not reprogrammed, serving only the application for which they were designed and manufactured. For these industry ASICs, operational CFP is the predominant contributor to the total CFP, followed by manufacturing and design CFP. The demonstration of how GreenFPGA can comprehensively model the CFP of industry testcases, highlighting each component based on input parameters of the testcase.

\section{Validation discussion and challenges}

It's crucial to emphasize that GreenFPGA serves as a tool for analyzing the embodied and operational CFP of FPGAs, facilitating comparative assessments against ASICs. This open-source methodology generates results based on the accuracy of input parameters. However, validating the output CFP values presents challenges due to the coarse nature of publicly available sustainability reports within the industry. These reports often aggregate CFP across all products for the year, encompassing unrelated downstream and upstream activities. Compounding the validation challenge, many input parameters, such as project durations and yields, are proprietary, making precise CFP measurements challenging. 
This work contributes by raising awareness through insights and relies on reasonable assumptions derived from data gathered from papers~\cite{act, imec-dtco, imec-wp, sudarshan2023ecochip} and reports~\cite{amd-report, TSMC-report, nvidia-report, apple-report, facebook-report}. Our comparative results between FPGAs and ASICs offer insights into their relative CFP differences (the absolute value of CFP is not the primary focus in this comparison). GreenFPGA is configurable with adjustable knobs for each input and assumption made which enhances its utility. This feature allows the tool to be employed by various users, including industrial architects, enabling sustainability-minded design decisions.
\section{Conclusion}
\label{sec:conclusion}
\noindent
We propose GreenFPGA, a tool to model CFP across the lifetime of FPGAs and help identify quantifiable scenarios under which FPGAs serve as sustainable acceleration alternatives to ASIC under iso-performance constraints. We demonstrated our experiments on different testcases and  found that FPGAs are greener alternatives for a large number of applications with low lifetime, and low volume.

\bibliographystyle{misc/ieeetr2}
\bibliography{misc/bibfile}

\balance
\flushend

\end{document}